\documentclass[%
reprint,
superscriptaddress,
amsmath,
amssymb,
aps,
prl,
]{revtex4-2}
\UseRawInputEncoding
\usepackage{graphicx}
\usepackage{subfigure}
\usepackage{dcolumn}
\usepackage{bm}
\usepackage{amsmath}
\usepackage{txfonts}
\usepackage[OT2,T1]{fontenc}
\usepackage[colorlinks=true,citecolor=blue, urlcolor = blue, linkcolor= blue]{hyperref}
\usepackage{epstopdf}
\usepackage{color,soul}
\usepackage[toc,page]{appendix}

\begin{document}

\title{Relaxation Oscillations of an Exciton-polariton Condensate Driven by Parametric Scattering}

\author{Chuan Tian}
\thanks{These authors contributed equally to this work.}
\affiliation{Wuhan National High Magnetic Field Center and School of Physics, Huazhong University of Science and Technology, Wuhan 430074, China}

\author{Linqi Chen}
\thanks{These authors contributed equally to this work.}
\affiliation{Key Laboratory of Materials for High-Power Laser, Shanghai Institute of Optics and Fine Mechanics, Chinese Academy of Science, Shanghai 201800, China}
\affiliation{Hangzhou Institute for Advanced Study, University of Chinese Academy of Sciences, Hangzhou 310024, China}

\author{Yingjun Zhang}
\affiliation{Wuhan National High Magnetic Field Center and School of Physics, Huazhong University of Science and Technology, Wuhan 430074, China}
\affiliation{Key Laboratory of Biomedical Engineering of Hainan Province, School of Biomedical Engineering, Hainan University, Haikou 570100, China}

\author{Liqing Zhu}
\affiliation{Wuhan National High Magnetic Field Center and School of Physics, Huazhong University of Science and Technology, Wuhan 430074, China}

\author{Wenping Hu}
\affiliation{Wuhan National High Magnetic Field Center and School of Physics, Huazhong University of Science and Technology, Wuhan 430074, China}

\author{Yichun Pan}
\affiliation{Wuhan National High Magnetic Field Center and School of Physics, Huazhong University of Science and Technology, Wuhan 430074, China}

\author{Zheng Wang}
\affiliation{Wuhan National High Magnetic Field Center and School of Physics, Huazhong University of Science and Technology, Wuhan 430074, China}

\author{Fangxin Zhang}
\affiliation{Wuhan National High Magnetic Field Center and School of Physics, Huazhong University of Science and Technology, Wuhan 430074, China}

\author{Long Zhang}
\affiliation{Key Laboratory of Materials for High-Power Laser, Shanghai Institute of Optics and Fine Mechanics, Chinese Academy of Science, Shanghai 201800, China}
\affiliation{Hangzhou Institute for Advanced Study, University of Chinese Academy of Sciences, Hangzhou 310024, China}

\author{Hongxing Dong}
\email{hongxingd@siom.ac.cn}
\affiliation{Key Laboratory of Materials for High-Power Laser, Shanghai Institute of Optics and Fine Mechanics, Chinese Academy of Science, Shanghai 201800, China}
\affiliation{Hangzhou Institute for Advanced Study, University of Chinese Academy of Sciences, Hangzhou 310024, China}

\author{Weihang Zhou}
\email{zhouweihang@hust.edu.cn}
\affiliation{Wuhan National High Magnetic Field Center and School of Physics, Huazhong University of Science and Technology, Wuhan 430074, China}
\date{\today}

\begin{abstract}
We report observation of coherent oscillations in the relaxation dynamics of an exciton-polariton condensate driven by parametric scattering processes. As a result of the interbranch scattering scheme and the nonlinear polariton-polariton interactions, such parametric scatterings exhibit high scattering efficiency, which leads to fast depletion of the polariton condensate and periodic shut-off of the bosonic stimulation processes, eventually causing relaxation oscillations. Employing polariton-reservoir interactions, the oscillation dynamics in the time domain can be projected onto the energy space. In theory, our simulations using the open-dissipative Gross-Pitaevskii equation are in excellent agreement with experimental observations. Surprisingly, the oscillation patterns are clearly visible in our time-integrated images including many excitation pulses, implying the high stability of the relaxation oscillations driven by polariton parametric scatterings. 
\end{abstract}

\maketitle
Exciton-polaritons, elementary excitations resulting from the strong coupling between semiconductor excitons and microcavity-confined photons, are attracting lots of attention in both fundamental and applied research. Being half-light half-matter in nature, exciton-polaritons represent an ideal platform for the study of a plethora of novel phenomena, such as solitons \cite{tanese2013polariton,walker2017dark,amo2011polariton}, non-Hermitian physics \cite{song2021room,liao2021experimental}, topological excitations \cite{klembt2018exciton} and quantum chaos \cite{gao2015observation}. As a result of their half-light nature, exciton-polaritons possess extremely small effective masses (typically on the order of $10^{-5}m_{e}$, with $m_{e}$ being free elelctron mass), leading to the ability of Bose Einstein condensation at temperatures readily accessible in laboratories \cite{su2020observation,dusel2020room,lerario2017room,zhang2020trapped,li2013from,kasprzak2006bose,sanvitto2016road}. Meanwhile, this half-light nature renders polaritons a short lifetime of a few picoseconds. They escape from the microcavity quickly and thus have to be replenished continuously. This nonequilibrium dissipative nature leads to complex nonlinear dynamics \cite{liu2020nonlinear,Galbiati2012polariton}. A solid understanding of polariton dynamics is thus critical for both fundamental description of light-matter condensation and their potential technological applications.

On the other hand, one of the most advantages of exciton-polaritons over purely photonic systems is the strong, and even tunable, particle interaction inherited from their half-matter nature \cite{tian2020polariton,sun2017direct,zhang2021van,estrecho2021low}. Because of this interaction, peculiar scattering processes, which are responsible for the nonlinearities, relaxation and thermalization of exciton-polaritons, can thus occur. Great effort has been devoted to the nonlinear polariton scattering processes, parametric scattering in particular \cite{savvidis2000angle,diederichs2006parametric,wu2021nonlinear,xie2012room,kuwata1997parametric,kundermann2003coherent}, in recent years. Compared with their optical counterpart, polariton parametric scatterings exhibit much higher parametric conversion efficiency \cite{diederichs2006parametric}, making them a promising candidate for the realization of on-chip micro-parametric oscillators. Moreover, polariton parametric scatterings also play a crucial role in many interesting macroscopic quantum phenomena, \textit{e.g.}, superfluidity \cite{Amo2009collective}, condensation \cite{xie2012room}, vortices \cite{sanvitto2010persistent,tosi2011onset}, \textit{etc}. Revealing the dynamics of polariton parametric scatterings and the physics behind is therefore an important step moving toward the proposed polaritronics.
\begin{figure}
	\centering
	{\includegraphics[width=0.48\textwidth] {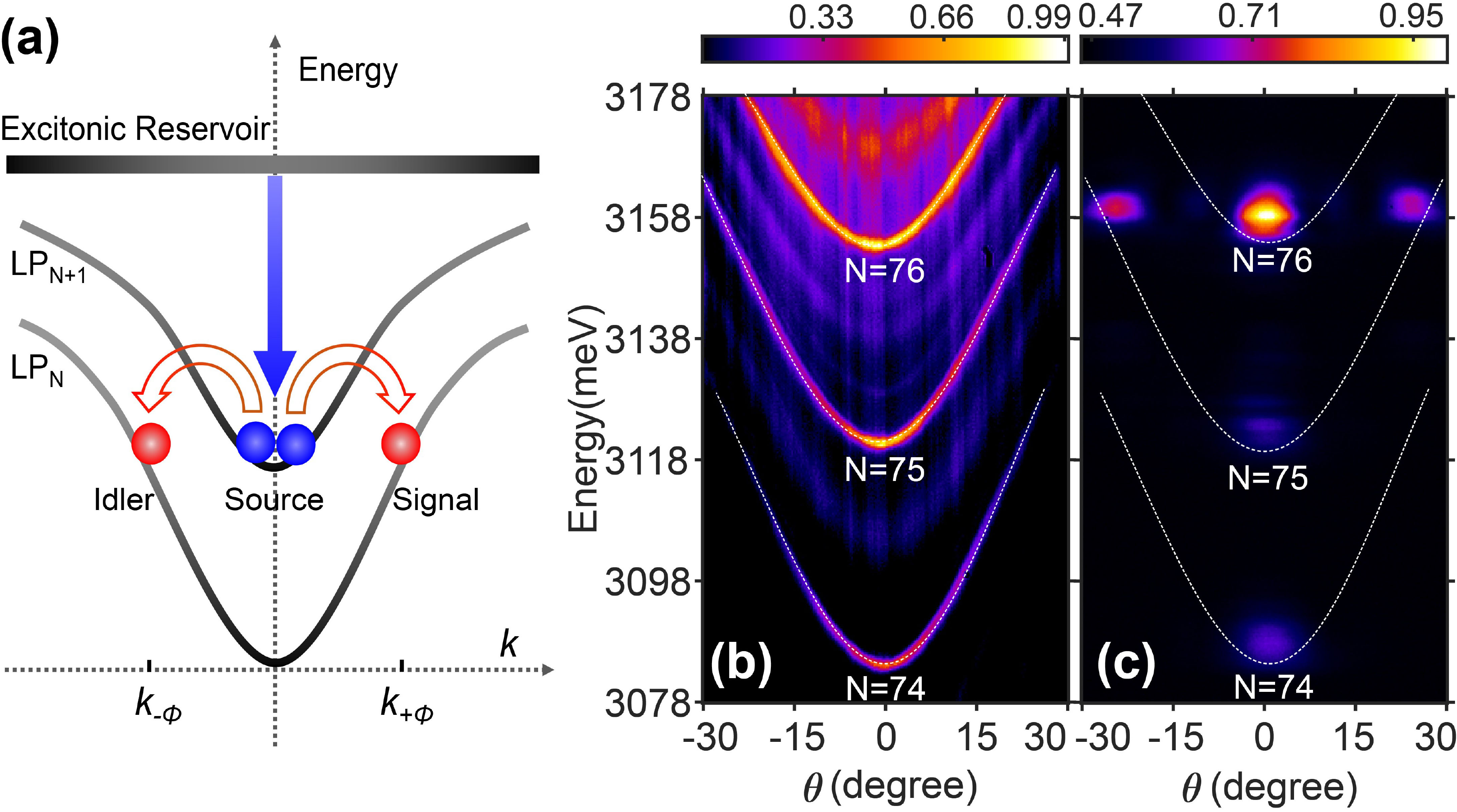}}
	\caption{(a) Schematics showing the mechanism of the relaxation oscillations of an exciton-polariton condensate driven by parametric scatterings. LP: lower polariton branch. (b) Typical angle-resolved micro-PL image for a ZnO microrod with radius of $\sim 2.28 \, \rm \mu m$ under below-threshold pumping using a 360 nm continuous-wave semiconductor laser. (c) Above-threshold angle-resolved micro-PL image for the same microrod excited by a 355 nm pulsed laser. Pumping power $P \approx 1.3 P_{th}$, with $P_{th}$ being the lasing threshold of the $76^{th}$ branch. White dotted curves in (b) and (c) are theoretical fittings using the coupled oscillator model. Photon content of the $76^{th}$ branch: $\sim 47\%$. }
	\label{fig1}
\end{figure}
\begin{figure*}
	\centering
	{\includegraphics[width=0.94\textwidth] {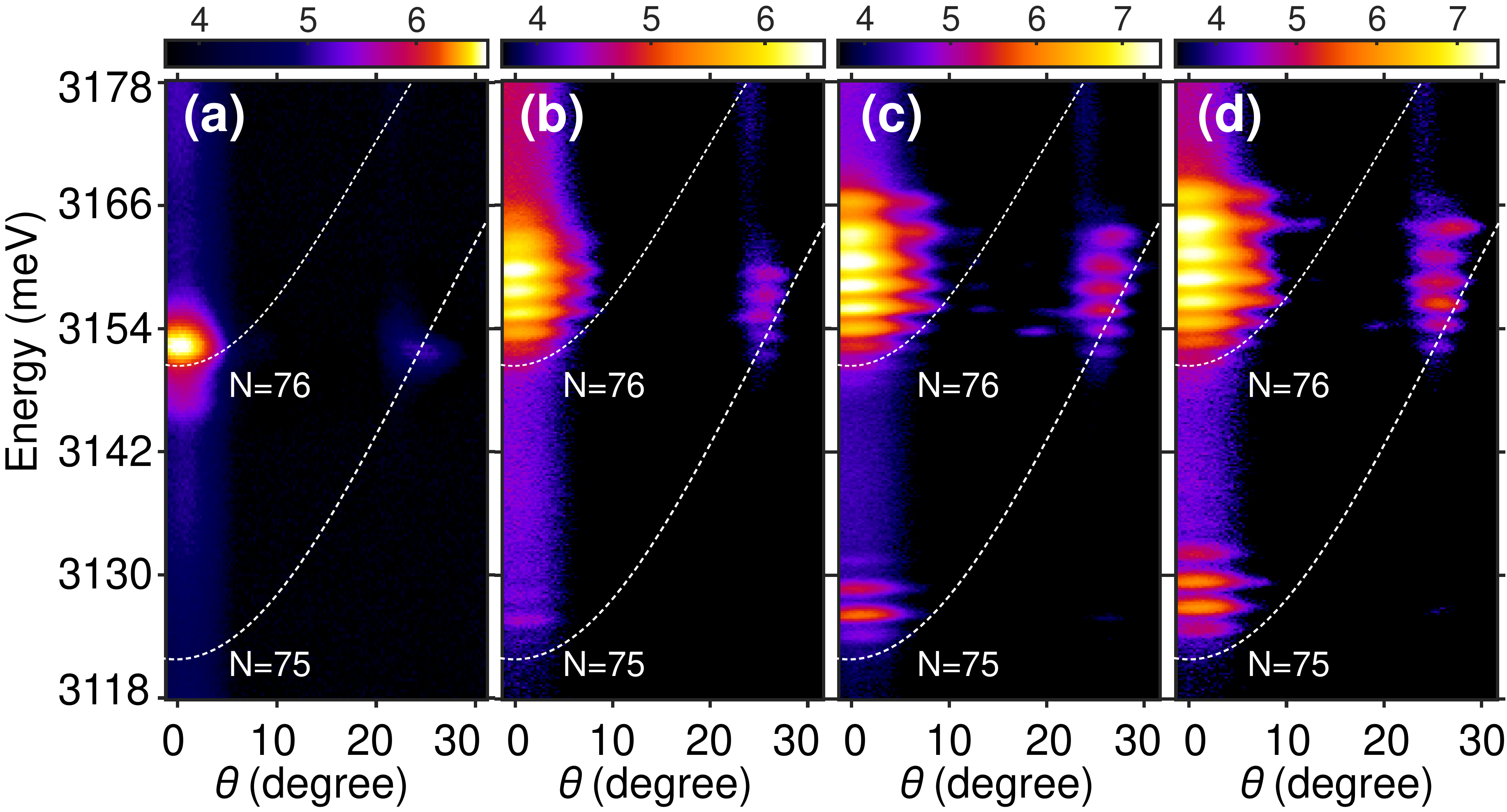}}
	\caption{ Zoomed-in angle-resolved micro-PL images showing the evolution of the source and signal beam patterns under different pumping powers. (a) Pumping power $P\approx1.1P_{th}$; (b) $P\approx1.9P_{th}$; (c) $P\approx2.5P_{th}$; (d) $P\approx2.8P_{th}$.}
	\label{fig2}
\end{figure*}

In this work, we report on the observation of relaxation oscillations of an exciton-polariton condensate driven by parametric scattering processes, a mechanism that has never been explored before. Commonly found in solid state lasers \cite{siegman1986lasers}, relaxation oscillations have been used to model a wide variety of nonlinear behaviors in many domains, such as electronics \cite{horowitz2015art}, electrochemistry \cite{hunt1990nonlinear} and biology \cite{kruse2005oscillations}. Mathematically, they are solutions to a certain class of coupled nonlinear differential equations and play an important role in understanding dynamics of complex systems. For nonequilibrium polariton condensates, spatial confinement \cite{giorgi2014relaxation} and the multi-stage relaxation involving both plasma and hot excitons \cite{horikiri2018transient,pieczarka2017relaxation} have been reported to be two possible mechanisms leading to relaxation oscillations. In contrast to these, we here show a completely different scenario that relaxation oscillations may arise from the interplay between parametric scatterings and Bose stimulation. The physical picture is as following: excitons relax from the reservoir via stimulated scattering and condense at the lowest state. Above threshold, parametric scatterings are triggered and quickly deplete the condensation, leading to the shut-off of the stimulated scattering until polaritons are replenished and the next cycle starts.

The samples we used are one-dimensional ZnO microrods grown by chemical vapor deposition method. Having smooth facets and hexagonal cross-sections, such microrods form natural whispering gallery (WG) microcavities for the confinement of light field \cite{zhang2018realization,tian2020polariton,xie2012room,luo2020classical}. The intrinsically large excitonic binding energy ($\sim{60}$ meV at room temperature) and oscillator strength make them a promising testing bed for experimental studies of polaritonics, as demonstrated in our previous work \cite{tian2020polariton,xie2012room,sun2008direct}. In our experiments, exciton-polaritons were created nonresonantly by a 360 nm continuous-wave semiconductor laser at low pumping power and a 355 nm pulsed laser (pulse width: 1 ns, repetition rate: 30 kHz) for intense pumping. The optical characteristics of exciton-polaritons were characterized by a home-built angle-resolved micro-photoluminescence (micro-PL) system equipped with a 40x ultraviolet objective lens (N.A.: 0.5) and a spectrometer with 550 mm focal length. All measurements were performed at room temperature.

Typical angle-resolved micro-PL image of a ZnO microrod at low pumping power is shown in Fig. 1(b). A series of parabola-like emission belts are clearly visible. These are the lower branches of exciton-polaritons resulting from the strong coupling between WG modes and ZnO excitons, as evidenced by the characteristic bending when approaching the exciton resonance at large angles (see Fig. S4(a)). The formation of multiple polariton branches is due to the relatively large size (radius: $\sim{2.28 \, \rm \mu m}$) of the WG cavity, which is a favorable situation for polariton parametric scatterings, as demonstrated in our previous work \cite{xie2012room}. In Fig. 1(c), we show the angle-resolved micro-PL image at pumping power slightly above lasing threshold ($\sim{1.3P_{th}}$), using a 355 nm pulsed laser. In strong contrast to the image in Fig. 1(b), all the three lower polariton branches shrink into bright spots at around $k_{//}=0$, showing the typical characteristics of polariton condensates. However, it should be noted that there are two additional bright spots symmetrically locating at around $\pm 25^{\circ}$ for the $N = 76$ branch. What's more, they are at the same energy position as the condensate spot at $0^{\circ}$ (equivalent to $k_{//}=0$). Obviously, these two additional spots come from the interbranch polariton parametric scatterings, as have been discussed thoroughly in our previous literature \cite{xie2012room}. In such process, two polaritons in the condensate interact with each other and are scattered to the next branch, generating a pair of balanced signal and idler beams with the same energy but opposite momenta. This scheme is shown schematically in Fig. 1(a). Compared with intrabranch parametric scatterings usually employed in two-dimensional planar microcavity systems \cite{savvidis2000angle}, the interbranch scattering scheme observed here has much higher scattering efficiencies, as conditions of momentum and energy conservation can be met more easily \cite{xie2012room}.
\begin{figure}
	\centering
	{\includegraphics[width=0.4\textwidth] {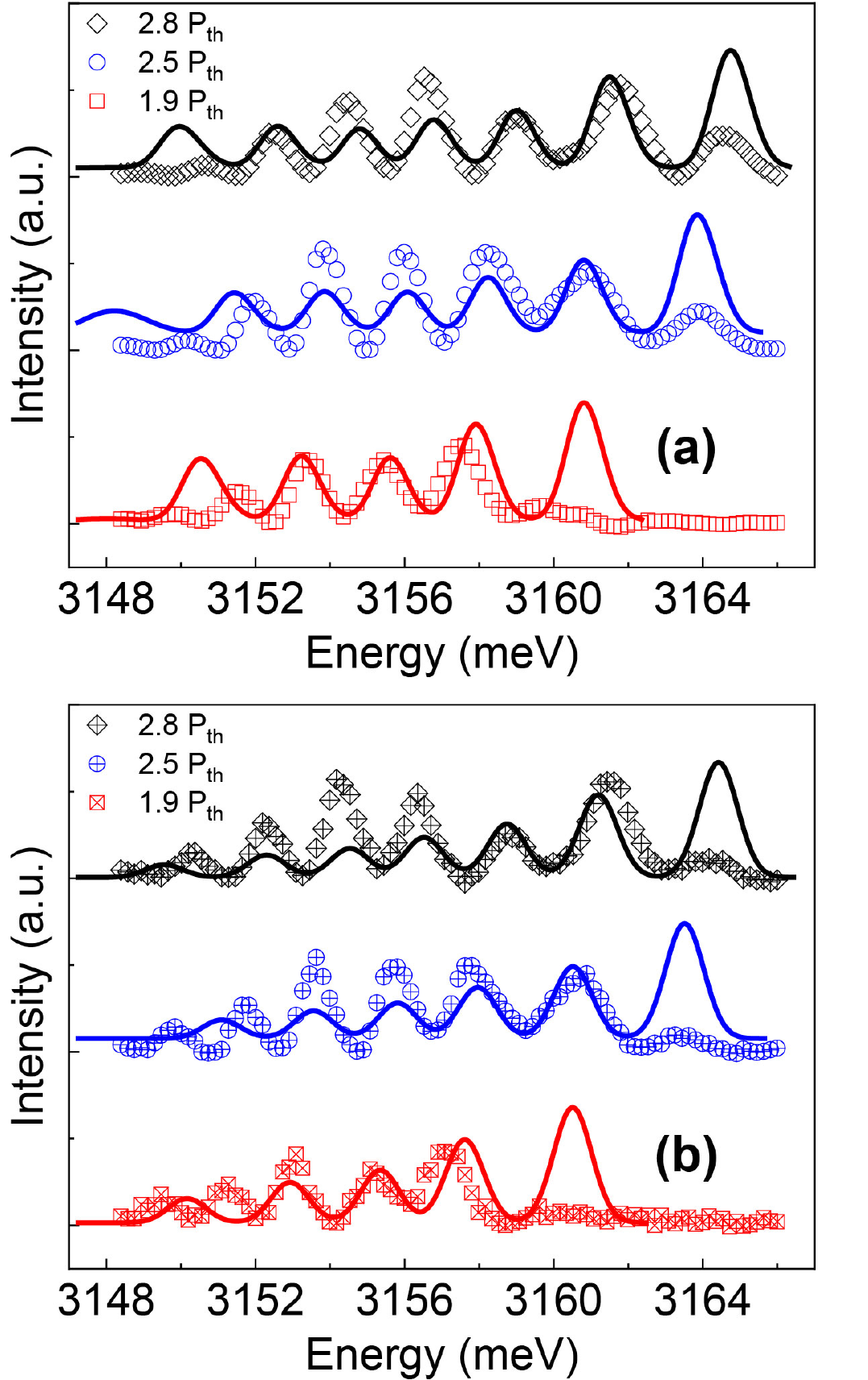}}
	\caption{ Spectral oscillations and theoretical simulations. (a) Spectra of the source beam at three representative pumping powers of $1.9P_{th}$ (red squares), $2.5P_{th}$ (blue circles) and $2.8P_{th}$ (black diamonds). (b) Corresponding spectra of the signal/idler beam at the same selected pumping powers of $1.9P_{th}$ (red filled squares), $2.5P_{th}$ (blue filled circles) and $2.8P_{th}$ (black filled diamonds). The solid curves in (a) and (b) are the corresponding simulated spectra using the Gross-Pitaevskii equations.}
	\label{fig3}
\end{figure}

\begin{figure*}
	\centering
	\includegraphics[width=1\textwidth]{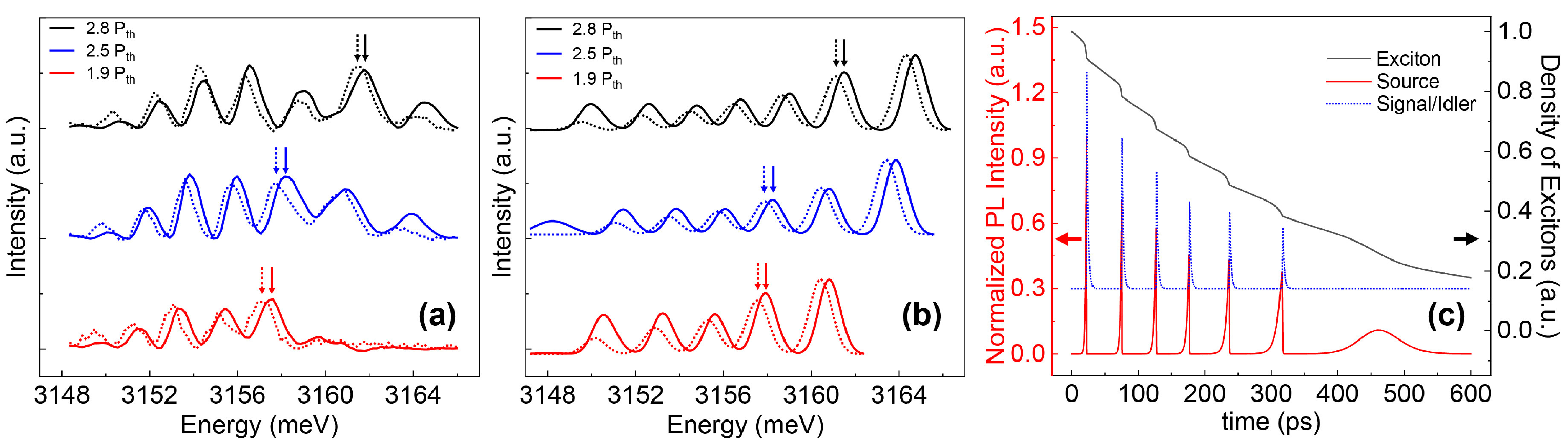}
	\caption{(a) Direct comparison between spectra of the source beams (solid curve) and their corresponding signal/idler beams (dotted curve) at three representative pumping powers of $1.9P_{th}$ (red), $2.5P_{th}$ (blue) and $2.8P_{th}$ (black). (b) Theoretically calculated spectra of the source (solid curve) and signal/idler (dotted curve) beams corresponding to those shown in (a). (c) Evolution of particle population in the time domain calculated using the Gross-Pitaevskii equation model at a pumping power of $2.5P_{th}$. Red solid (blue dotted) curve represents polaritons in the source (signal/idler) beam. Black solid curve denotes the population evolution of excitons in the reservoir. The curve for signal/idler beam has been shifted vertically for clarity. Relevant parameters: $m = 2.0\times10^{-5}m_{e}$, $\it \Gamma \rm =7.0\times10^{-3} \, \rm \mu m\cdot \rm ps^{-1}$, $\gamma _{R} = 0.002 \, \rm ps^{-1}$, $\gamma_{0} = 0.2 \, \rm ps^{-1}$, $\gamma_{\varPhi} = 0.5\, \rm ps^{-1}$, $\gamma_{P} = 0$, $\beta_0 = 1.5\times10^{-5}  \, \rm \mu m\cdot \rm ps^{-1}$, $\beta_{\varPhi}= 0$, $g_{p} = 0.25\,\rm\mu eV\cdot\, \rm \mu m$, $g_{R} = 0.50 \,\rm\mu eV\cdot\, \rm \mu m$, FWHM of laser pulse = $2.0 \, \rm \mu m$.}
	\label{fig4}
\end{figure*}

For a deeper understanding of polariton parametric scatterings, we performed power-dependent analyses. Typical results are shown in Fig. 2. For clarity, we show the zoomed-in images which cover the $75^{th}$ and $76^{th}$ branches in angular region [$-2^{\circ}$, $30^{\circ}$] only. Fig. 2(a) is the image at pumping power slightly above lasing threshold ($\sim{1.1P_{th}}$), where conventional parametric scattering patterns are again clearly visible. However, drastic changes were found when pumping power was increased to $\sim{1.9P_{th}}$. As one can see from Fig. 2(b), the bright spot of polariton condensate now turns into interference-like fringes. With further increase of the pumping power, more stripes, along with significant blueshift, can be identified unambiguously, as demonstrated in Fig. 2(c) and 2(d). The signal beam, on the other hand, shows very similar behaviors in its power dependence. To further check their characteristics, we convert the observed fringes into conventional spectra for both source and signal/idler beams. The resulting spectra corresponding to those images shown in Fig. 2(b-d) are plotted in Fig. 3(a) and (b) for the source and signal beams, respectively. As demonstrated, oscillation-like spectral modulations are again identified unambiguously.

Having revealed the striking spectral features, we now turn to their physical origin. One possibility is that these fringes are interference patterns resulting from the phase coherence of polariton condensates. Suppose there are two or more traps created by defects or impurities in the pumping area, the trapped condensates, which are spatially separated but phase-coupled, may indeed lead to interference fringes. However, these fringes change little as we move the detection spot along the microrod (Fig. S5, Supplementary Information). This contradicts the general knowledge that traps induced by defects or impurities are strongly position-sensitive. Therefore, this possibility can be ruled out. Considering the fact that such fringes develop from the parametric scattering patterns, it is quite natural to conjecture that the observed fringes are closely related to polariton parametric scatterings. Keeping in mind that polaritons here are injected nonresonantly, we propose a mechanism involving both excitonic reservoir and polariton condensate. The physical picture of our proposal is shown schematically in Fig.1(a): following the creation of an excitonic reservoir by the external laser, hot excitons relax (via phonon emission and polariton scattering) and accumulate on the lowest state of the $(N+1)^{th}$ branch. Above threshold, the accumulation of polaritons triggers the stimulated scattering process, leading to the formation of a polariton condensate. The establishment of the condensate, on the other hand, initiates the interbranch polariton parametric scatterings, which quickly deplete the condensate and shut off the stimulated scattering process until the next cycle. As the excitonic reservoir is decaying in time, the blueshift of the polariton states, which depends on the exciton density, is also decreasing. The relaxation oscillations in the time domain will therefore be projected onto the energy space.

To verify our conjecture, we carried out theoretical simulations using the well-known Gross-Pitaevskii equation which has been proven to be very successful for the description of polariton condensates \cite{xue2014creation,tanese2013polariton,luo2020classical,haug2014quantum}. The equations describing our proposed model can be written as following


\begin{eqnarray}
		i\hbar\dfrac{\partial\varPsi_{0}}{\partial t}=&& \left[-\dfrac{\hbar^{2}}{2m}\dfrac{\partial^{2}\varPsi_{0}}{\partial x^{2}}+g_{p}(\lvert\varPsi_{0}\lvert^{2}+\lvert\varPsi_{\varPhi}\lvert^{2})+g_{R}N_{R}\right. \nonumber\\
     	&&\left.+\dfrac{i\hbar}{2}(\beta_{0}N_{R}-\gamma_{0}-\it \Gamma\lvert\varPsi_{\varPhi} \rvert^{2}) \right]\varPsi_{0}
\end{eqnarray}
\begin{eqnarray}
	i\hbar\dfrac{\partial\varPsi_{\varPhi}}{\partial t}=&& \left[-\dfrac{\hbar^{2}}{2m}\dfrac{\partial^{2}\varPsi_{\varPhi}}{\partial x^{2}}+g_{p}(\lvert\varPsi_{0}\lvert^{2}+\lvert\varPsi_{\varPhi}\lvert^{2})+g_{R}N_{R}\right. \nonumber\\
	&&\left.+\dfrac{i\hbar}{2}(\beta_{\varPhi}N_{R}-\gamma_{\varPhi}+\it \Gamma\lvert\varPsi_{0} \rvert^{2}) \right]\varPsi_{\varPhi}
\end{eqnarray}
\begin{eqnarray}
\dfrac{\partial N_{R}}{\partial t}=&&-\gamma_{R}N_{R}-\beta_{0}N_{R}\lvert \varPsi_{0} \rvert^{2}-\beta_{\varPhi}N_{R}\lvert \varPsi_{\varPhi} \rvert^{2}\nonumber\\
&&+P(x)e^{-\gamma_{P}t}
\end{eqnarray}Here, evolution of the source condensate, signal (idler) beams and the excitonic reservoir is described by Eqs. (1-3), respectively. $\varPsi_{0}$ ($\varPsi_{\varPhi}$) is wavefunction of the source (signal/idler) beam, $N_{R}$ represents the population of excitons in the reservoir, $\gamma_{0}$ ($\gamma_{\varPhi}$) is the decay rate of the source (signal/idler) beam, $\gamma_{R}$ is the decay rate of the reservoir, $\beta_{0}$ ($\beta_{\varPhi}$) means the stimulated scattering rate between reservoir and the source (signal/idler) beam, $\it \Gamma$ refers to the rate of polariton parametric scatterings, $P(x)$ is a Gaussian function describing the spatial profile of the laser pulse and $\gamma_{P}$ represents its decay rate. $g_{p}$ ($g_{R}$), on the other hand, stands for the mean field polariton-polariton (polariton-reservoir) interaction constant. Here, it should be noted that the interaction constant $g_{p}$ was taken to be the same for both source and signal/idler polaritons, as the differences between them are rather small. Typical simulation results for the time evolution of particle population at pumping power of $2.5P_{th}$ are plotted in Fig. 4(c). Interestingly, with reasonable parameters, both source condensate and signal/idler beam show marked oscillations of their population with period around 60 ps. The exciton population, which shows a monotonic decrease, was also given (black curve), for the readers' convenience. Noticeably, the exciton curve exhibits a step-like characteristics, instead of a smooth curve. This is essentially a manifestation of the bosonic stimulation which causes a fast consumption of the excitonic reservoir. To obtain the oscillation patterns in the energy space, we employ the mean field approximation, where blueshift of the polariton states can be expressed as $\Delta E = g_{p}·n_{p} + g_{R}·n_{R}$ ($n_{p}$: polariton density, $n_{R}$: exciton density) \cite{tian2020polariton}. Considering the fact that contributions from polariton-polariton interactions are far smaller than those from polariton-reservoir interactions under non-resonant optical pumping condition, the pure polariton-polariton interaction term was neglected in the calculation of $\Delta E$. By doing so, we can readily obtain the relaxation oscillation modulated spectra for the source and signal/idler beams, as shown by the solid curves in Fig. 3(a) and (b), respectively. Fantastically, the simulated spectra are in very good agreement with our experimental results, thus supporting our proposal that the observed fringes stem from relaxation oscillations driven by polariton parametric scatterings. 

Besides the spectral modulations, there are also other footprints showing relaxation oscillations of polaritons. As plotted in Fig. 4(c), we notice that the spikes of the signal/idler beams (blue dotted curve) always lag behind their counterparts of the source beam (red solid curve). As the excitonic reservoir is decaying, this lag will surely cause redshift of the signal/idler spikes in the energy domain relative to those of the source beam. To verify this prediction, we stack the spectra of both source and signal/idler beams together, as shown in Fig. 4(a) for three representative pumping powers of $1.9P_{th}$, $2.5P_{th}$ and $2.8P_{th}$. Expectedly, peaks of the signal/idler spectra do show significant redshift compared with their source beam counterparts for all selected pumping powers. This supports again that the fringes in the angle-resolved images result from polariton relaxation oscillations driven by parametric scatterings. As a comparison, the corresponding simulated spectra are given in Fig. 4(b), where excellent agreements can be found with our experimental results.

In addition to the phenomena discussed above, it is noteworthy that the lower lying $75^{th}$ branch also exhibits similar patterns, as can be seen clearly from Fig. 2. This is actually quite expectable considering the multi-branch polariton structures studied in our work. However, differences can also be found when checked carefully. Compared with those patterns in the $76^{th}$ branch, the most significant difference is that the number of stripes is much smaller than its counterpart. This difference, however, can be well rationalized by taking the lower excitonic fraction of polaritons in the $75^{th}$ branch into account. Indeed, as parametric scatterings among polaritons are essentially governed by the excitonic constituent of polaritons, the reduced excitonic fractions will surely lead to lower parametric scattering efficiencies. What's more, reduction in excitonic fraction will lead to longer lifetime of polaritons, thus inducing smaller loss rate and change the characteristics of relaxation oscillations. To unveil the effects of excitonic fraction on polariton relaxation oscillations, we carried out simulations again using the same model described by Eqs. (1-3). By varying the excitonic fraction but keeping pumping power unchanged, our simulations do confirm that polaritons with lower excitonic fraction exhibit fewer cycles of oscillations in their relaxation dynamics, as shown in supplementary Fig. S7. Satisfactory agreements can also be found between patterns of the $75^{th}$ branch and our simulations, as demonstrated by Fig. S8.

In summary, we revealed coherent oscillations in the relaxation dynamics of an exciton-polariton condensate injected by non-resonant optical pumping. Combining angle-resolved power-dependent measurements with Gross-Pitaevskii equation modeling, we verified that the relaxation oscillations observed in this work are driven by polariton parametric scattering processes. This is a new mechanism for relaxation oscillations that has never been explored before. Through polariton-reservoir interactions, the oscillations in the time domain were successfully transferred to the energy space, providing an alternative, with greatly reduced complexity, for the studies of polariton dynamics. The balanced signal and idler pulses created by relaxation oscillations observed here also open a new way for the generation of entangled photon pulses with coherence. Our findings could thus find technological applications in polariton-based quantum devices.

\begin{acknowledgments}
This work was financially supported by National Natural Science Foundation of China (NSFC) (Grant No. 52172140) and the Interdisciplinary Program of Wuhan National High Magnetic Field Center (Grant No. WHMFC202111), Huazhong University of Science and Technology; the Fundamental Research Funds for the Central Universities, HUST (Grant No. 2019kfyXKJC005).
\end{acknowledgments}

\end{document}